\documentclass[a4paper,english,10pt,superscriptaddress,reprint,amsmath,amssymb,aps,twocolumn,floatfix,notitlepage,showkeys,prd]{revtex4-2}
\usepackage{graphicx}
\usepackage{bm}
\usepackage{dcolumn}
\usepackage[colorlinks=true,citecolor=darkred,urlcolor=darkred, pdfborder={0 0 0}]{hyperref}
\usepackage{setspace}
\usepackage{caption}
\usepackage{subcaption} 
\usepackage{slashed}
\usepackage[normalem]{ulem}
\usepackage{float}
\usepackage[usenames,dvipsnames]{xcolor}  
\usepackage{microtype}
\usepackage{lipsum}
\usepackage{amsmath}
\usepackage{amssymb}
\usepackage{mathrsfs}
\usepackage{placeins}
\usepackage{lettrine}   
\input Eileen.fd        
\newcommand*\initfamily{\usefont{U}{Eileen}{xl}{n}}

\definecolor{darkred}{rgb}{0.6,0,0}
\definecolor{linkcolor}{rgb}{0,0,0.5}
\usepackage[T1]{fontenc} 
\usepackage[compat=1.1.0]{tikz-feynhand}
\usepackage{tikz-feynman}
\tikzfeynmanset{compat=1.1.0}
\usepackage{feynmp}
\usepackage{tikzsymbols}
\usepackage{array}
\usepackage{pifont} 

\definecolor{linkcolor}{rgb}{0,0,0.5}
\usepackage{orcidlink}
\begin{document}
\preprint{2408.xxxx}
\title{\boldmath \color{BrickRed}Non-Holomorphic Modular $\mathcal{A}_4$ Symmetric Scotogenic Model}
%
\author{Takaaki Nomura \orcidlink{0000-0002-0864-8333}}
\email{nomura@scu.edu.cn}
\affiliation{College of Physics, Sichuan University, Chengdu 610065, China}
\author{Hiroshi Okada \orcidlink{0000-0002-4573-1822}}
\email{okada@hnu.edu.cn}
\email{okada.hiroshi@phys.kyushu-u.ac.jp}
\affiliation{Department of Physics, Henan Normal University, Henan 453007, China}
\author{Oleg Popov \orcidlink{0000-0002-0249-8493}}
\email{opopo001@ucr.edu (corresponding author)}
\affiliation{Faculty of Physics and Mathematics, Shenzhen MSU-BIT University,\\ 1, International University Park Road, Shenzhen 518172, China}
\date{\today}
%
\begin{abstract}
The present work extends scotogenic and its modular $\mathcal{A}_4$ variation a step forward and demonstrates scotogenic modular $\mathcal{A}_4$ non-supersymmetric realization. To achieve this non-holomorphic modular symmetries come to rescue. Advantage of the current construction is the compactness of the model content and absence of the supersymmetric fields. Neutrino mass is generated through a canonical scotogenic mechanism. The allowed values of the VEV of the $\tau$ modulus are $\tau \simeq w$ and Im$[\tau]\approx 2$. The non-holomorphic modular $\mathcal{A}_4$ symmetry leads to correlations among the neutrino observables.
\end{abstract}
\maketitle 
%
\section{Introduction}
\label{sec:intro}
%
\lettrine[lines=4,findent=-1.3cm]{\normalfont\initfamily \fontsize{17mm}{10mm}\selectfont A \normalfont\initfamily}{ } Radiative seesaw model proposed by Ernest Ma~\cite{Ma:2006km} is among the most attractive scenarios to derive naturally tiny neutrino masses at loop level in addition to direct connection of a dark matter candidate. Furthermore, the model generates sizable lepton flavor violating processes (LFVs) that would be in the reach of current experiments. It suggests that the scotogenic mechanism does work in the framework of a non-supersymmetric theory (non-SUSY).

On the other hand, a modular flavor group is such a powerful symmetry which provides numerous predictions on neutrino masses and their mixing patterns~\cite{Feruglio:2017spp}. Further, it plays a role in stabilizing the dark matter candidate by appropriately assigning the modular weights to particles~\cite{Nomura:2019jxj}. 
However, the original modular group would require holomorphicity in order NOT to induce infinite non-suppressed terms that spoil predictabilities of models. 
This forces model builders to work within SUSY paradigm whenever one aims to utilize modular flavor symmetry in the context of the neutrino mass model.

Recently, "Qu" and "Ding" have mathematically shown that modular symmetries are consistent with a non-holomorphic theory~\cite{Qu:2024rns}. That is a breakthrough paper for "AUTHENTIC" phenomenologist because one can apply modular flavor symmetry to any non-SUSY model!~\footnote{After the original paper, this symmetry has been applied to the flavor physics; $A_4$ in~\cite{Nomura:2024atp} $S_4$ in \cite{Ding:2024inn}.} 

In this letter, we revisit the modular $\mathcal{A}_4$ radiative seesaw model~\cite{Nomura:2019lnr} and apply non-holomorphic modular $A_4$ symmetry. Then, we construct a minimum model and show several predictions for both cases of normal and inverted hierarchies.

The paper is organized as follows: Sec.~\ref{sec:model} describes the model and its content, neutrino mass generation is given in Sec.~\ref{sec:m_nu}, constraints on the model's parameters are presented in Sec.~\ref{sec:constraints}, results are given in Sec.~\ref{sec:numerical}, Sec.~\ref{sec:discussion} contains discussion and the summary.
\section{Model}
\label{sec:model}
\begin{table}[h]
    \centering
    \begin{tabular}{cccccc}
        \hline \hline
        Fields & $SU(3)_c$ & $SU(2)_L$ & $U(1)_Y$ & $\mathcal{A}_4$ & $k$ \\ \hline
        $\overline{L_L}$ & $\pmb{1}$ & $\pmb{2}$ & $\frac{1}{2}$ & $\pmb{3}$ & -1 \\
        $e_R$ & $\pmb{1}$ & $\pmb{1}$ & $-1$ & $\pmb{3}$ & 1 \\
        $N_R$ & $\pmb{1}$ & $\pmb{1}$ & $0$ & $\pmb{3}$ & 0 \\ \hline
        $H$ & $\pmb{1}$ & $\pmb{2}$ & $~\frac{1}{2}$ & $\pmb{1}$ & 0 \\ 
        $\eta$ & $\pmb{1}$ & $\pmb{2}$ & $-\frac{1}{2}$ & $\pmb{1}$ & 1 \\ 
        \hline \hline
    \end{tabular}
    \caption{Model field content.}
    \label{tab:model_fields}
\end{table}
In this section, we summarize set up of the model based on non-holomorphic modular $A_4$ symmetry.
Due to non-holomorphic nature, we do not need to adopt SUSY framework. Then field contents are the same as original scotogenic model introducing SM singlet fermion $N_R$ and inert doublet scalar $\eta$.
We assign $A_4$ triplet to leptons including $N_R$ while the other fields are the trivial $A_4$ singlet. 
Also non-zero modular weight is assigned to lepton doublet $\overline{L_L}$, charged lepton singlet $e_R$ and  $\eta$  as $-1$, $1$ and $1$, respectively. The field contents and charge assignments are summarized in Table~\ref{tab:model_fields}.
Notice that we do not need $Z_2$ symmetry to forbid undesired terms $\overline{N_R} L_L H$ and $H \eta$ since they are not allowed by modular $A_4$ symmetry due to our choice of modular weight.  

The Lagrangian for lepton sector and non-trivial term of scalar potential are represented by
\begin{subequations}
    \label{eq:lag}
    \begin{align}
    \label{eq:lepton}
\mathcal{L}_{Y_\ell} & = M_0 [\overline{N^c_R} N_R]_{\pmb{1}}
+ M_3 Y_{\pmb{3}}^{(0)} [\overline{N^c_R} N_R]_{\pmb{3}_s} \\
&+ y_\ell [\overline{L_L} e_R]_{\pmb{1}} H + y_{\ell_3} Y_{\pmb{3}}^{(0)} [\overline{L_L} e_R]_{\pmb{3}_s} H \nonumber \\
&+ y'_{\ell_3} Y_{\pmb{3}}^{(0)} [\overline{L_L} e_R]_{\pmb{3}_a} H 
+ y_D [\overline{L_L} N_R]_{\pmb{1}} \eta \nonumber \\
& + y_{D_3} Y_{\pmb{3}}^{(0)} [\overline{L_L} N_R]_{\pmb{3}_s} \eta
+ y'_{D_3} Y_{\pmb{3}}^{(0)} [\overline{L_L} N_R]_{\pmb{3}_a} \eta \nonumber \\
& +\text{h.c.}, \nonumber
\end{align}
\end{subequations}
\setcounter{equation}{0} 
\begin{subequations}
     \setcounter{equation}{1}
    \label{eq:lag_v}
    \begin{align}
        V & = \mu_\eta^2  \eta^\dagger \eta + \tilde{\lambda} Y_{\pmb{1}}^{(-2)} (H \eta)^2  + \lambda_{H\eta} \left(H^\dagger H\right)\left(\eta^\dagger \eta\right) \nonumber \\
&  + \lambda_{H\eta}^\prime \left|H \eta\right|^2 + \text{other trivial terms},
    \end{align}
\end{subequations}

where the notation $[\cdots]_{\pmb{r}}$ indicates $A_4$ representation of $\pmb{r}$ is constructed inside the square bracket, $\pmb{3}_{s(a)}$ is $A_4$ triplet symmetrically(asymmetrically) constructed from two triplets, 
and $Y^{(k)}_{\pmb{r}}$ denotes the modular form with $A_4$ representation $\pmb{r}$ and modular weight $k$.
Here we write $Y_{\pmb{3}}^{(0)} = (y_1, y_2, y_3)$ where the component $y_{1,2,3}$ is given by Maa{\ss} forms~\cite{Qu:2024rns}. 
The Yukawa interactions are invariant under the modular $A_4$ when the corresponding operators are $A_4$ singlet and have zero modular weight. Also $\eta^\dagger \eta$ is invariant by the same way as kinetic term where we absorbed some factors including modulus in free parameters.

The inert scalar doublet $\eta$ is represented by
\begin{equation}
\eta = \begin{pmatrix} \frac{1}{\sqrt2} (\eta_R + i \eta_I) \\ \eta^- \end{pmatrix},
\end{equation}
where $\eta^-$ has electric charge $-1$.
After electroweak symmetry breaking, we obtain squared masses of inert scalar bosons such that
\begin{align}
m_{\eta_{R[I]}}^2 &= \mu^2_\eta + \frac{v^2}{2}(\lambda_{H \eta} + \lambda'_{H \eta} +[-] 2 \lambda), \nonumber \\
m_{\eta^\pm}^2 &= \mu^2_\eta + \frac{v^2 \lambda_{H \eta}}{2},
\end{align}
where $\lambda \equiv \tilde{\lambda} Y_{\pmb{1}}^{(-2)}$ and $v$ is the vacuum expectation value (VEV) of the SM Higgs field.
Thus squared mass difference between $\eta_R$ and $\eta_I$ is given by $\Delta m^2 \equiv m_{\eta_R}^2 - m_{\eta_I}^2 = 2 \lambda v^2$. 
In numerical analysis below we consider $\Delta m^2$ is free parameter instead of $\lambda$ and assume $m_{\eta_I} \simeq m_{\eta^\pm}$ for simplicity.

Applying $A_4$ multiplication rules, we obtain Majorana mass matrix for $N_R$ as follows
\begin{align}
M_N&= M_3
\left[\begin{array}{ccc}
\tilde{M}_0+ 2y_{1}  & -y_{3}  & -y_{2}  \\ 
-y_{3} & 2y_{2} & \tilde{M}_0-y_{1} \\ 
-y_{2} & \tilde{M}_0-y_{1} &2 y_{3} \\ 
\end{array}\right],
\end{align}
where $\tilde{M}_0 \equiv M_0/M_3$.
The Majorana mass matrix is diagonalized by a unitary matrix $U_N$ as $D_N\equiv U_N^T M_N U_N$.

The Yukawa interaction among $\overline {L_L}$, $N_R$ and $\eta$ is written by the form of $(\overline {L_L} y_\eta N_R) \eta$. The Yukawa matrix $y_\eta$ is explicitly represented by
{\small
\begin{align}
& y_\eta = \nonumber \\
& \left[\begin{array}{ccc}
y_D+ 2 y_{D_3} y_{1}  & -(y_{D_3}-y'_{D_3})y_{3}  & -(y_{D_3}+y'_{D_3})y_{2}  \\ 
-(y_{D_3}+y'_{D_3})y_{3} & 2 y_{D_3} y_{2} & y_D-(y_{D_3}-y'_{D_3})y_{1} \\ 
-(y_{D_3}-y'_{D_3})y_{2} & y_D -(y_{D_3}+y'_{D_3})y_{1} &2 y_{D_3} y_{3} \\ 
\end{array}\right].
\end{align}
}
After electroweak symmetry breaking, the charged lepton mass matrix is obtained such that
\begin{align}
& \frac{m_\ell}v
= \nonumber \\
& \left[\begin{array}{ccc}
y_\ell+ 2 y_{\ell_3} y_{1}  & -(y_{\ell_3}-y'_{\ell_3})y_{3}  & -(y_{\ell_3}+y'_{\ell_3})y_{2}  \\ 
-(y_{\ell_3}+y'_{\ell_3})y_{3} & 2 y_{\ell_3} y_{2} & y_\ell-(y_{\ell_3}-y'_{\ell_3})y_{1} \\ 
-(y_{\ell_3}-y'_{\ell_3})y_{2} & y_\ell -(y_{\ell_3}+y'_{\ell_3})y_{1} &2 y_{\ell_3} y_{3} \\ 
\end{array}\right].
\end{align}
%
The mass matrix is diagonalized with unitary matrices $V_R, V_L$ as diag.$(m_e,m_\mu,m_\tau)\equiv V_R^\dag m_\ell V_L$.
We need $V_L$ matrix in our numerical analysis that is obtained by the relation $V_L^\dag m_\ell^\dag m_\ell V_L ={\rm diag.}(|m_e|^2,|m_\mu|^2,|m_\tau|^2)$.
In addition, we fix the free parameters $y_\ell,y_{\ell_3},y'_{\ell_3}$ to fit the three observed charged-lepton masses numerically, applying the following three relations:
\begin{subequations}
\label{eq:ml_rel}
\begin{align}
& {\rm Tr}[m_\ell^\dag m_\ell] = |m_e|^2 + |m_\mu|^2 + |m_\tau|^2, \\
& {\rm Det}[m_\ell^\dag m_\ell] = |m_e|^2  |m_\mu|^2  |m_\tau|^2, \\
& \left( {\rm Tr}\left[m_\ell^\dag m_\ell \right] \right)^2 - {\rm Tr} \left[ \left( m_e^\dag m_e \right)^2 \right] \nonumber \\
& = 2 \left( |m_e|^2  |m_\mu|^2 + |m_\mu|^2  |m_\tau|^2+ |m_e|^2  |m_\tau|^2 \right). \label{eq:l-cond}
\end{align}
\end{subequations}

\vspace{-1cm}
\section{Neutrino Mass}
\label{sec:m_nu}
The active Majorana neutrino mass terms are induced at one-loop level where $N$ and $\eta$ propagate the inside loop diagram.
The mass matrix is approximately derived as 
\begin{align}
&m_{\nu_{ij}}\approx \sum_{\alpha=1-3}\frac{Y_{\eta_{i\alpha}} {D_{N}}_\alpha Y^T_{\eta_{\alpha j}}}{(4\pi)^2} \\
& \times
\left(\frac{m_R^2}{m_R^2-{ D^2_N}_{\alpha}}\ln\left[\frac{m_R^2}{{ D^2_N}_{\alpha}}\right]
-
\frac{m_I^2}{m_I^2-{D^2_N}_{\alpha}}\ln\left[\frac{m_I^2}{{ D^2_N}_{\alpha}}\right]
\right), \nonumber
\end{align}
where $Y_\eta = y_\eta U_N$. The neutrino mass matrix is diagonalized by a unitary matrix $V_\nu$ as $D_\nu \equiv V_\nu^T m_\nu V_\nu$ where $D_\nu$ is diagonalized one ($D_\nu = \{m_1, m_2, m_3 \}$). The Pontecorvo-Maki-Nakagawa-Sakata (PMNS) mixing matrix, $U (\equiv U_{\rm PMNS})$, is defined as $U = V_L^\dag V_\nu$ since charged lepton mass matrix is not diagonal in flavor basis. 
Mixing angles can be estimated in terms of the component of $U$ such that
\begin{subequations}
\begin{align}
& \sin^2\theta_{13}=|U_{13}|^2, \\
&\sin^2\theta_{23}=\frac{|U_{23}|^2}{1-|U_{13}|^2}, \\
& \sin^2\theta_{12}=\frac{|U_{12}|^2}{1-|U_{13}|^2}.
\end{align}
\end{subequations}
Dirac CP phase $\delta_{CP}$ is derived from PMNS matrix elements, computing the Jarlskog invariant, as follows
\begin{equation}
J_{CP} = \text{Im} [U_{11} U_{2 2} U_{1 2}^* U_{2 1}^*] = s_{23} c_{23} s_{12} c_{12} s_{13} c^2_{13} \sin \delta_{CP},
\end{equation}
where $s_{ij}(c_{ij})$ stands for $\sin \theta_{ij} (\cos \theta_{ij})$.
The Majorana phases are also estimated in terms of the other invariant quantities $I_1$ and $I_2$:
\begin{align}
I_1 &= \text{Im}[U^*_{11} U_{12}] = c_{12} s_{12} c_{13}^2 \sin \left( \frac{\alpha_{21}}{2} \right), \nonumber \\
I_2 &= \text{Im}[U^*_{11} U_{13}] = c_{12} s_{13} c_{13} \sin \left( \frac{\alpha_{31}}{2} - \delta_{CP} \right).
\end{align}
In addition, we evaluate 
the effective mass for the neutrinoless double beta by
\begin{align}
m_{ee}=\left| m_{1} c^2_{12} c^2_{13}+m_{2} s^2_{12} c^2_{13}e^{i\alpha_{21}}
+m_{3} s^2_{13}e^{i(\alpha_{31}-2\delta_{CP})} \right|.
\end{align}
Its value is tested by experiments, e.g. KamLAND-Zen~\cite{KamLAND-Zen:2016pfg}. 

In numerical analysis, we rewrite neutrino mass matrices as
\begin{equation}
m_\nu(m_R, m_I, D_{N_\alpha}) = M_3 \tilde{m}_\nu (\tilde{m}_R, \tilde{m}_I, \tilde{D}_{N_\alpha}),
\end{equation}
where $\tilde{m}_\nu$ is dimensionless matrix with $\tilde{m}_{R,I} \equiv m_{R,I}/M_3$ and $\tilde{D}_{N_\alpha} \equiv D_{N_\alpha}/M_3$.
The eigenvalues of $\tilde{m}_\nu$ are also denoted by $\{\tilde{m}_1,\tilde{m}_2,\tilde{m}_3 \}$
In this way, overall factor $M_3$ can be estimated as 
\begin{align}
{\rm(NH):} M_3^2 = \frac{\Delta m_{\rm atm}^2}{\tilde{m}_3^2 - \tilde{m}_1^2}, \quad
{\rm(IH):} M_3^2 = \frac{\Delta m_{\rm atm}^2}{\tilde{m}_2^2 - \tilde{m}_3^2}, \quad
\end{align}
where $\Delta m_{\rm atm}^2$ is atmospheric neutrino mass square difference, and NH and IH indicate the normal and  inverted hierarchy, respectively.
\vspace{-1cm}
\section{Constraints}
\label{sec:constraints}
In this section we summarize phenomenological constraints on the model.

The sum of neutrino mass is constrained by cosmological observations. The analysis with the Planck CMB data~\cite{Planck:2018vyg} provides $\sum m_\nu \lesssim 120$ meV (95\% C.~L.) under the standard $\Lambda$CDM cosmological model.
In addition more stringent constraint is obtained as $\sum m_\nu < 72$ meV (95\% C.~L.) if we include recent baryon acoustic oscillation (BAO) analysis by Dark Energy Spectroscopic Instrument (DESI) data~\cite{DESI:2024mwx}; even stronger bounds are estimated in refs.~\cite{Craig:2024tky, Wang:2024hen}. 
In numerical analysis we discuss these constraints denoted by "CMB" and "CMB+BAO" 

{\it  Charged lepton flavor violating (cLFV) processes} are induced through Yukawa interactions associated with $y_\eta$ coupling where we need to take into account mixing via $V_L$ and $U_N$ matrices to calculate the processes in mass basis.
Considering these mixing, we obtain the BRs such that~\cite{Baek:2016kud}
\begin{align}
&{\rm BR}(\ell_i\to\ell_j\gamma) \nonumber \\ 
& \approx
\frac{48\pi^3\alpha_{em}C_{ij}}{G_F^2 (4\pi)^4}
\left|\sum_{\alpha=1-3}Y'_{\eta_{j\alpha}} Y'^\dag_{\eta_{\alpha i}} F(D_{N_\alpha},m_{\eta^\pm})\right|^2,
\end{align}
\begin{align}
&F(m_a,m_b) \nonumber \\ & \approx \frac{2 m^6_a+3m^4_am^2_b-6m^2_am^4_b+m^6_b+12m^4_am^2_b\ln\left(\frac{m_b}{m_a}\right)}{12(m^2_a-m^2_b)^4},
\end{align}
where $Y'_\eta\equiv V_L^\dag y_\eta U_N$, $C_{21}=1$, $C_{31}=0.1784$, $C_{32}=0.1736$, $\alpha_{em}(m_Z)=1/128.9$, and $G_F=1.166\times10^{-5}$ GeV$^{-2}$.
Currently experimental upper bounds of the BRs are given 
by~\cite{MEG:2016leq, BaBar:2009hkt,Renga:2018fpd}
\begin{subequations}
\label{eq:lfvs-cond}
\begin{align}
& {\rm BR}(\mu\to e\gamma)\lesssim 3.1\times10^{-13}, \\
& {\rm BR}(\tau\to e\gamma)\lesssim 3.3\times10^{-8}, \\
& {\rm BR}(\tau\to\mu\gamma)\lesssim 4.4\times10^{-8}.
\end{align}
\end{subequations}

Inert charged scalar boson mass is also constrained from collider search since it could be produced via electroweak process if it is light enough. We adopt the constraints by LEP
experiment~\cite{ALEPH:2013htx} as $m_{\eta^\pm} \gtrsim 100$ GeV in numerical analysis.

\section{Numerical Analysis}
\label{sec:numerical}

In this section we show results of numerical analysis.
We globally scan free parameters of in the model within the following region:
\begin{align}
& {\rm Im}[\tau] < 10, \  \{ \tilde{M}_0, \tilde{m}_R \} \in [10^{-2}, 10^2], \
\Delta m^2 \in [10^{-5}, 1], \nonumber \\
& \{|y_D|, |y_{D_3}|, |y'_{D_3}| \} \in [10^{-5}, 1],
\end{align}
where $\Delta \tilde{m}^2 \equiv \Delta m^2/M^2_3$.
The LFV constraints are imposed in the analysis and we also require $m_{\eta^\pm} > 100$ GeV to avoid constraint from the LEP experiment.
We then search for the parameters that can fit neutrino data of NuFit 5.2~\cite{Esteban:2020cvm}, and calculate observables as predictions.

\begin{figure}[tb]
\begin{center}
\includegraphics[width=40.0mm]{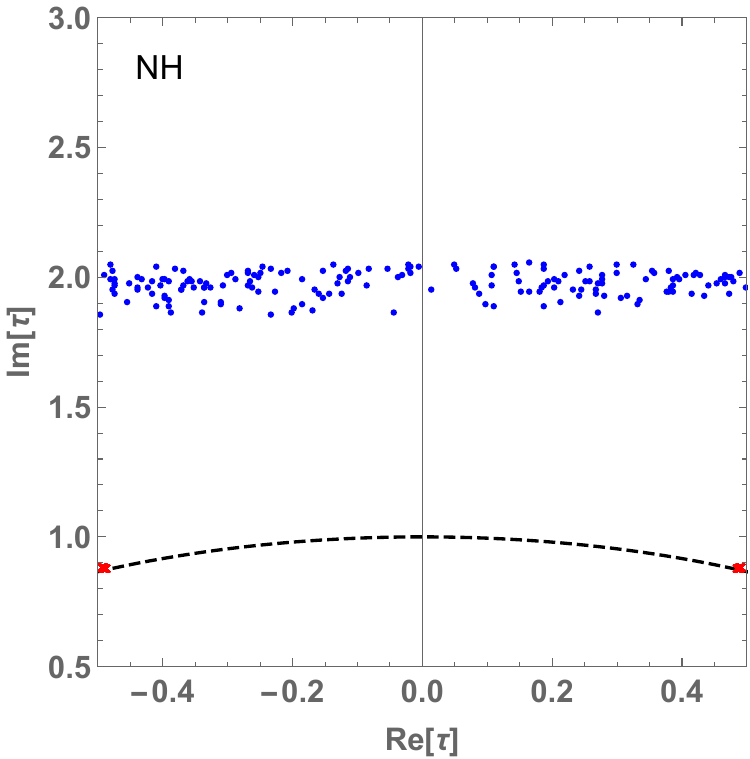} \quad
\includegraphics[width=40.0mm]{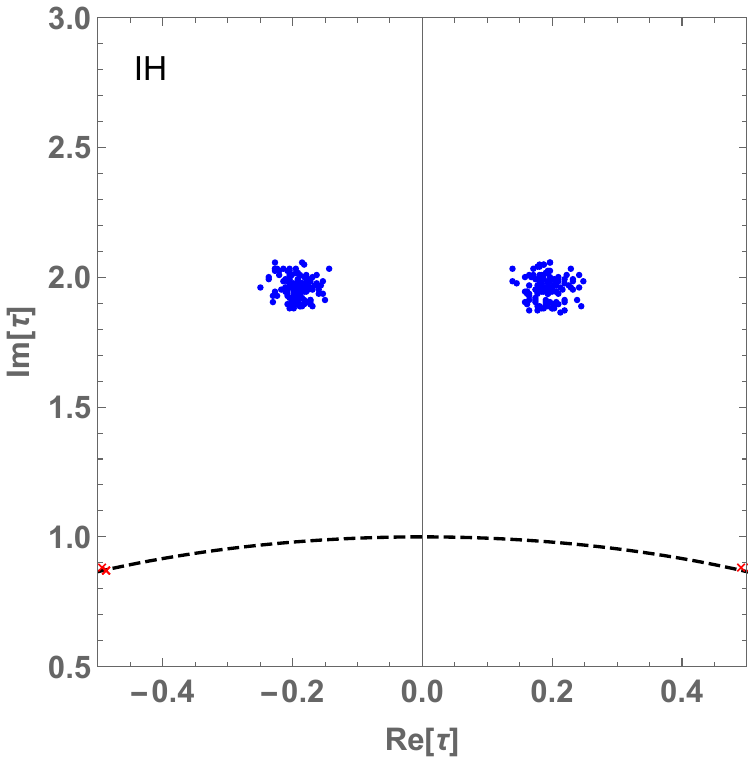} \quad
\caption{Allowed points on complex $\tau$ plane. The points near the fixed point $\tau = \omega$ are distinguished by red cross while the other region is given by blue points. The red cross and blue points are used in the same way in the other figures below.  }
  \label{fig:tau}
\end{center}\end{figure}

Fig.~\ref{fig:tau} shows the allowed values of modulus $\tau$. Remarkably we find the allowed points near the fixed point $\tau = \omega (=e^{2 \pi i/3})$ in both NH and IH case where these points are represented by red $"\times"$ mark to distinguish from other points. We also find allowed region around Im$[\tau] \sim 2$ where Re$[\tau]$ is also localized around $\pm 0.2$ in IH case but NH case does not show any tendency for Re$[\tau]$. Note that the red cross and blue points are used in the same way in the other figures below.

Fig.~\ref{fig:angle1} shows correlation between $\delta_{CP}$ and $\alpha_{21}$ estimated by allowed parameters.
Some correlations are found in both NH and IH cases. In particular these angles are localized in $\{|\delta_{CP}|,|\alpha_{31]}| \} \in \{[20,40], [80,100] \}$ degree  near the fixed point for IH case.

\begin{figure}[tb]
\begin{center}
\includegraphics[width=40.0mm]{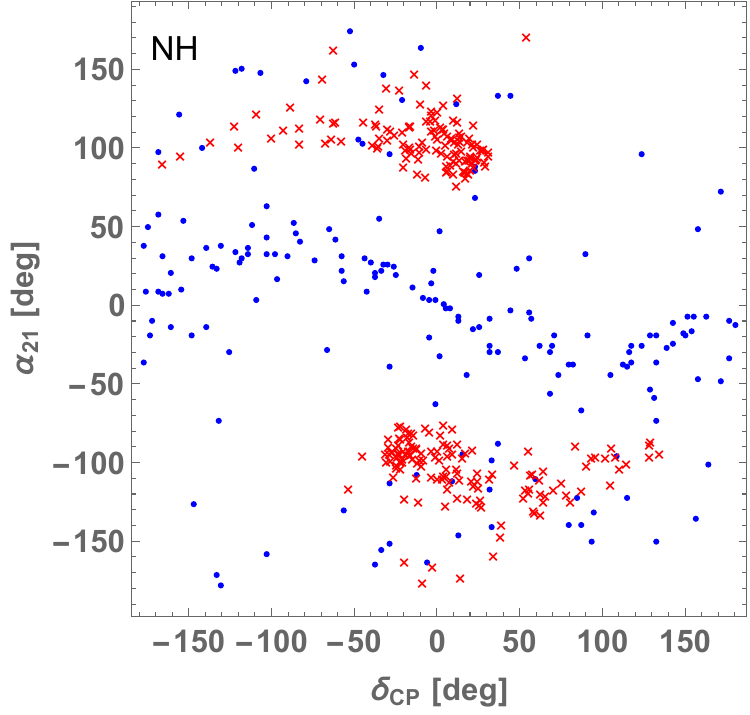} \quad
\includegraphics[width=40.0mm]{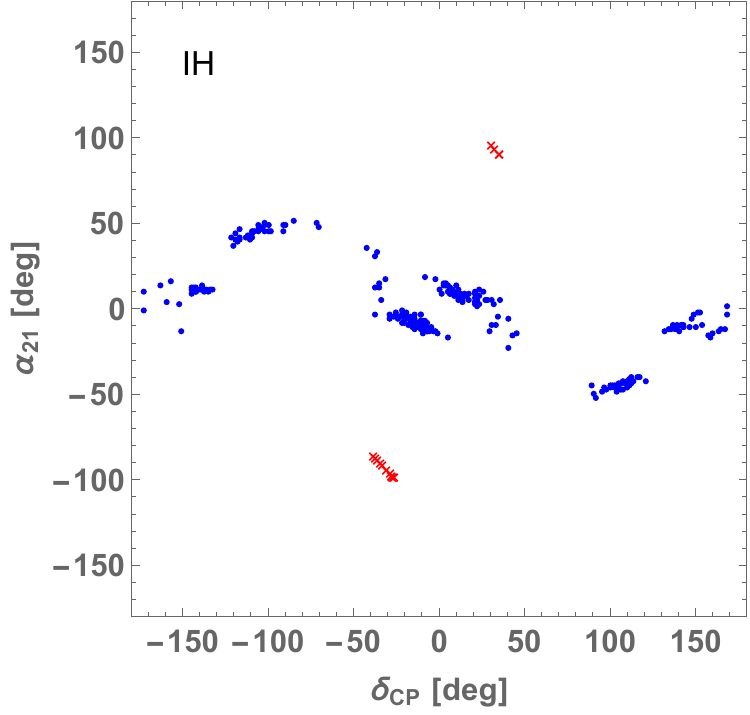} \quad
\caption{Values of $\delta_{\rm CP}$ and $\alpha_{21}$ estimated from allowed parameters.}
  \label{fig:angle1}
\end{center}\end{figure}

\begin{figure}[tb]
\begin{center}
\includegraphics[width=40.0mm]{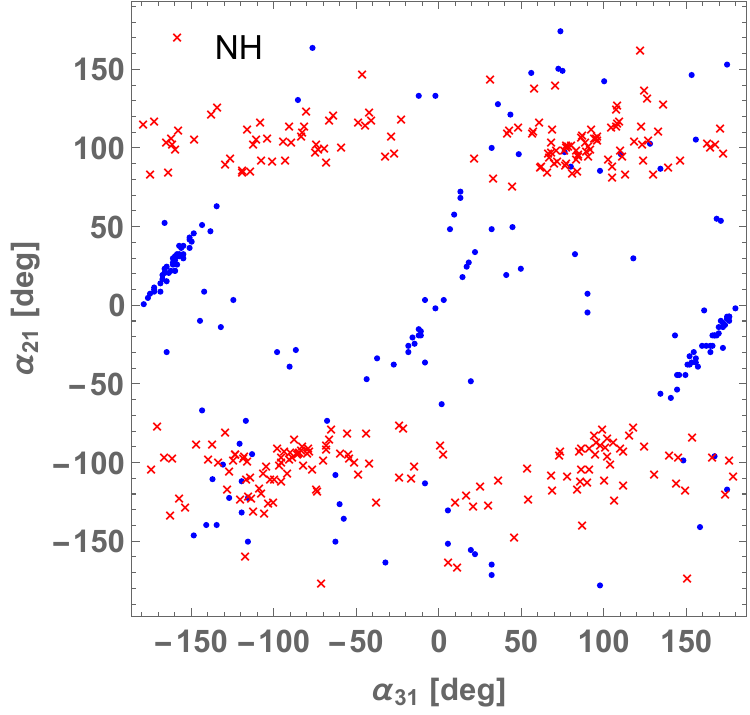} \quad
\includegraphics[width=40.0mm]{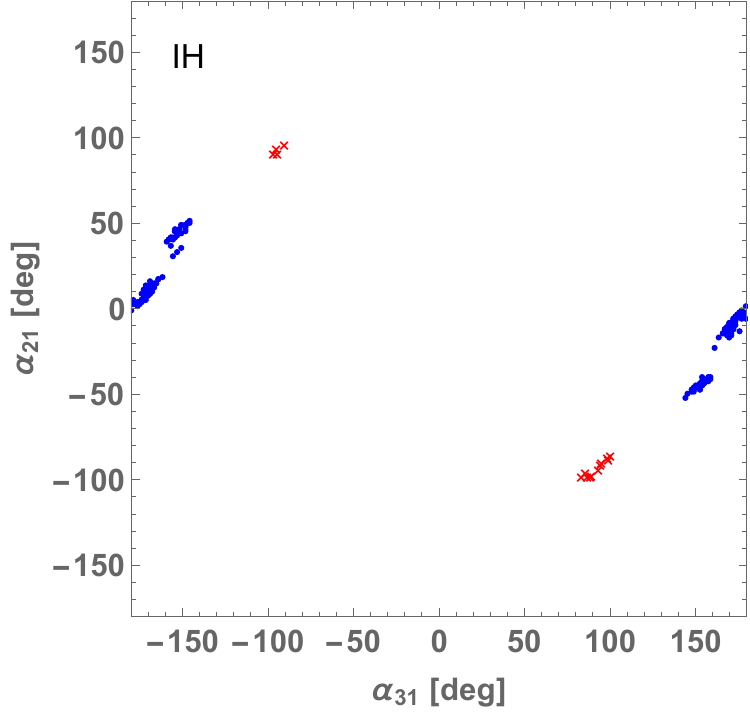} \quad
\caption{Values of $\alpha_{31}$ and $\alpha_{21}$ estimated from allowed parameters.}
  \label{fig:angle2}
\end{center}\end{figure}

In Fig.~\ref{fig:angle2} we show correlation between two Majorana phases for allowed parameters.
For NH case, $\alpha_{21}$ value tends to be within $[\pm 70, \pm 170]$ for the fixed point while wider region is covered in other $\tau$ point; some regions are more preferred containing lager number of points. For IH case, we obtain more limited region of Majorana phases where angle $\alpha_{21(31)}$ tends to be around $\pm 100 (\mp 100)$ degree for $\tau \simeq \omega$.
Other region of $\tau$ also give us limited $\alpha_{21(31)}$ within $[-50, 50]([\pm 150, \pm 180])$ degree.

\begin{figure}[tb]
\begin{center}
\includegraphics[width=40.0mm]{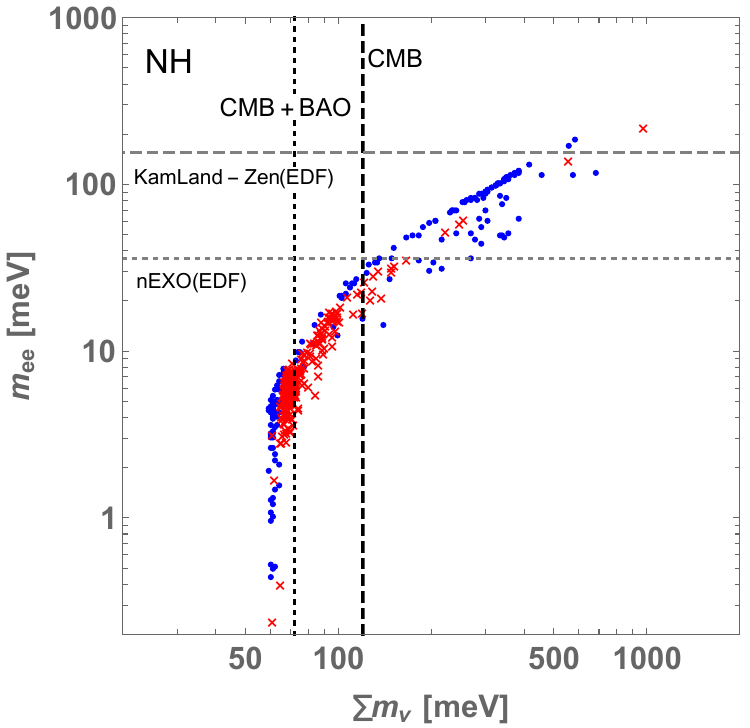} \quad
\includegraphics[width=40.0mm]{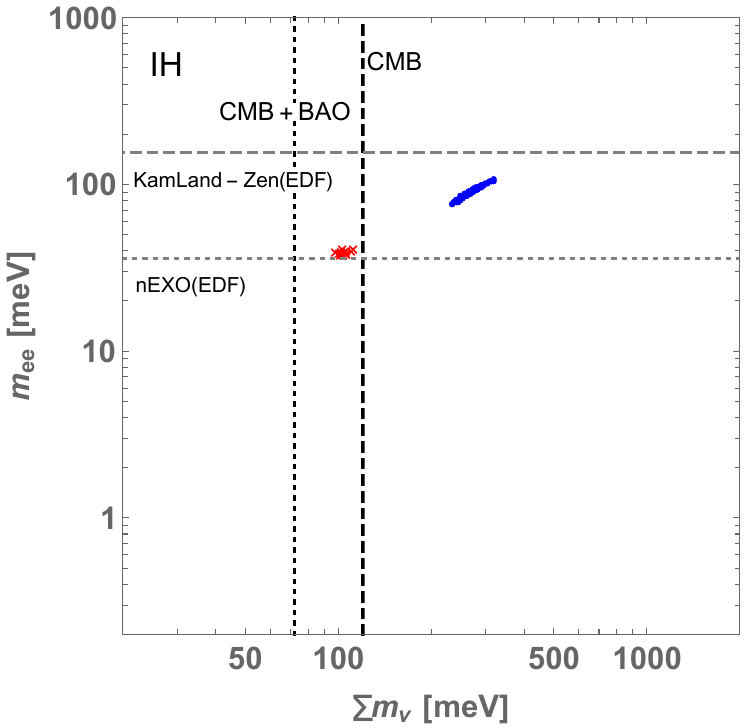} \quad
\caption{Values of $\sum m_\nu$ and $m_{ee}$ estimated from allowed parameters.}
  \label{fig:mass1}
\end{center}\end{figure}

The predicted values of $\sim m_\nu$ and $m_{ee}$ are shown in Fig.~\ref{fig:mass1}. The dashed and dotted vertical lines show the cosmological bound on $\sum m_\nu$ by the Planck CMB data and by combination of the CMB and the DESI BAO data respectively. In addition, the dashed and dotted horizontal lines respectively indicate the current constraint from KamLand-Zen~\cite{KamLAND-Zen:2022tow} and future prospect in nEXO~\cite{nEXO:2017nam} with energy-density functional (EDF) theory for nuclear matrix element.
For NH case, we have points satisfying both the $\sum m_\nu$ limit. On the other hand, for IH case, we only have points satisfying $\sum m_\nu$ limit from CMB only for $\tau \simeq \omega$. Thus the fixed point is preferred in IH case from cosmological bound on $\sum m_\nu$.

\begin{figure}[tb]
\begin{center}
\includegraphics[width=40.0mm]{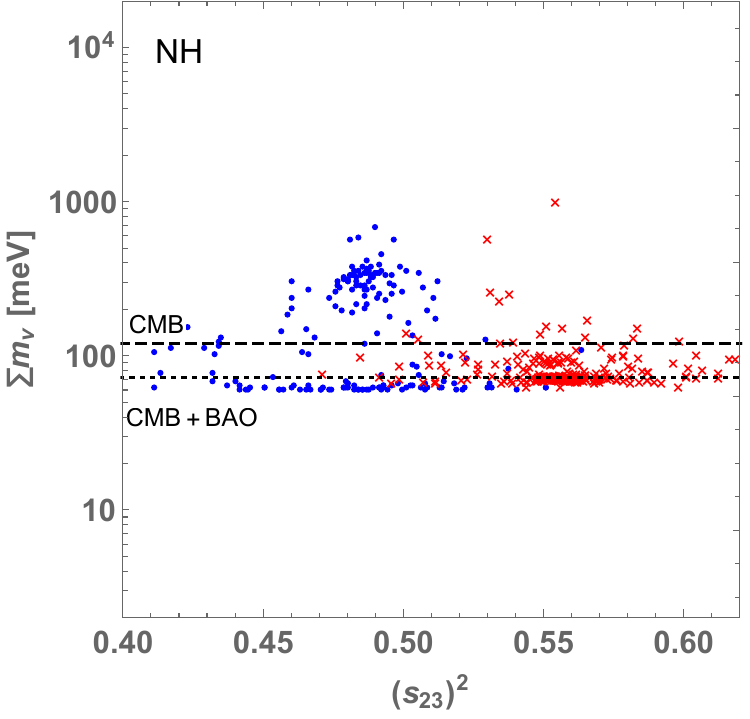} \quad
\includegraphics[width=40.0mm]{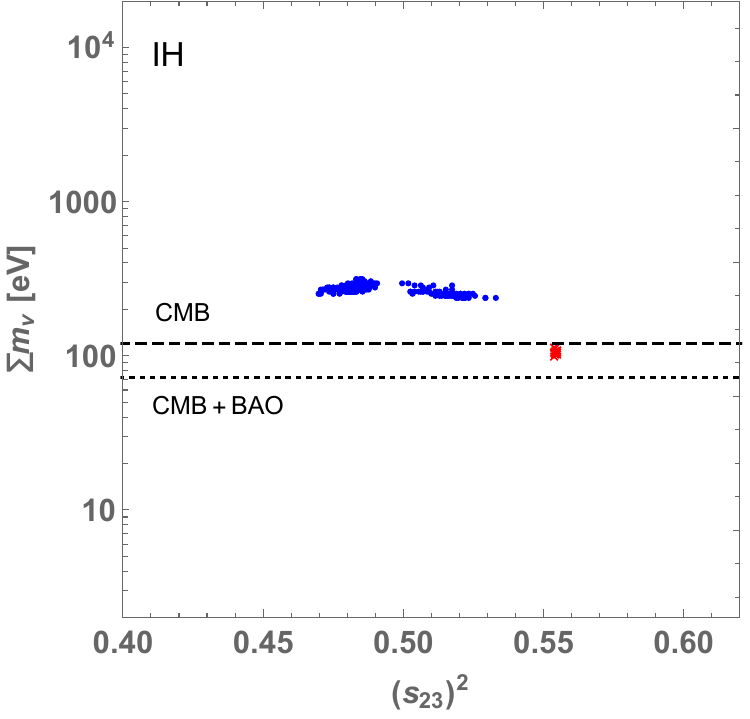} \quad
\caption{Values of $s_{23}^2$ and $\sum m_\nu$ estimated from allowed parameters.}
  \label{fig:mass2}
\end{center}\end{figure}

In Fig.~\ref{fig:mass2}, we show allowed points on $\{s^2_{23},\sum m_\nu \}$ plane. Remarkably, $s^2_{23}$ tend to be larger for fixed point of $\tau$ compared to other point of $\tau$. In particular the fixed point in IH case indicates $s^2_{23}$ is within $[0.55, 0.56]$ that is close to the best fit value.

\begin{figure}[tb]
\begin{center}
\includegraphics[width=40.0mm]{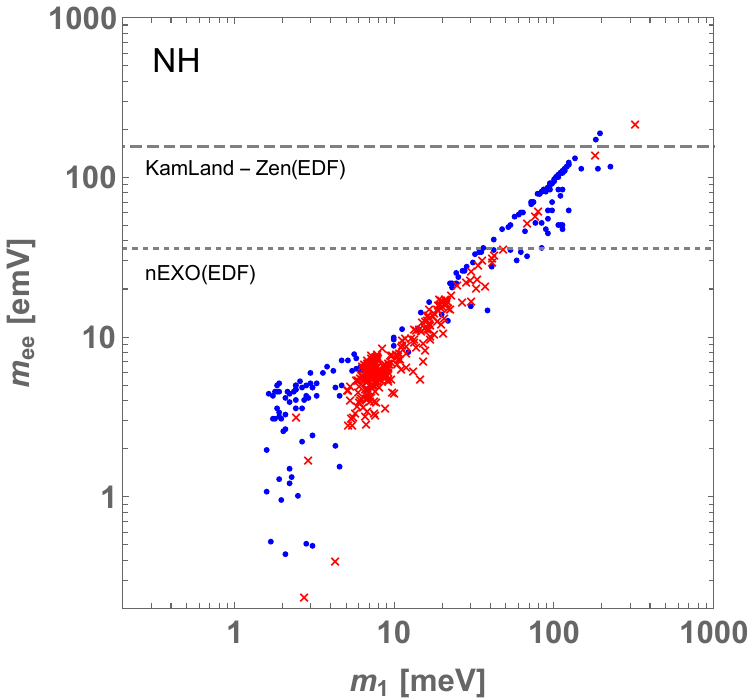} \quad
\includegraphics[width=40.0mm]{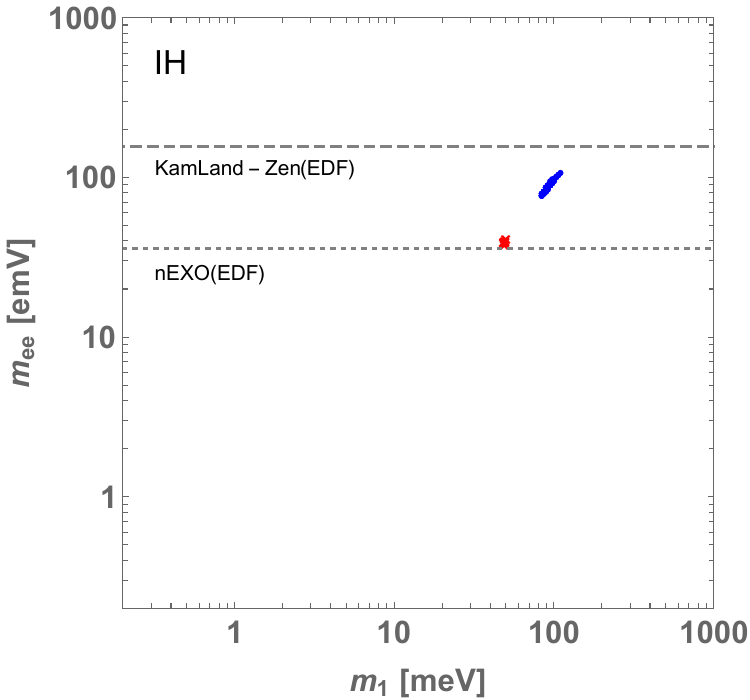} \quad
\caption{Values of $m_1$ and $m_{ee}$ estimated from allowed parameters.}
  \label{fig:mass3}
\end{center}\end{figure}

Finally, Fig.~\ref{fig:mass3} shows predicted points on $\{m_1, m_{ee} \}$ plane. We also find very limited region in IH case. In particular, the lightest neutrino mass is predicted to be $m_1 \sim 50$ meV for $\tau \simeq \omega$.

%
\section{Summary and discussion}
\label{sec:discussion}
We have studied a scotogenic model with non-holormorphic modular flavor $A_4$ symmetry.
The minimal field contents of the original scotogenic model can be directly applied since we do not need to impose supersymmetry.
The lepton fields including singlet $N$ have been all assigned as $A_4$ triplet universally. 
We have then constructed Lagrangian which is invariant under the gauge and flavor symmetries and derived charged lepton mass matrix and Majorana mass matrix of $N$.

The neutrino mass matrix has been derived at loop level via Yukawa interactions among lepton doublet, inert Higgs doublet and singlet fermion, $N$, where its structure is constrained by the modular $A_4$ symmetry.
We have then carried out numerical analysis searching for allowed parameters accommodating neutrino data to obtain predictions of the model taking phenomenological constraints into account.
In both NH and IH cases, allowed region of $\tau$ has been found near the fixed point $\tau \simeq \omega$ and Im$[\tau] \sim 2$.
We have also found some correlations among CP violating phases; especially the values of them are limited in IH case.
For neutrino mass sum, NH case can satisfy both CMB and CMB+BAO constraints while IH case can only satisfy CMB constraint near the fixed point of $\tau$.
Also, all the allowed region of IH case can be tested by future neutrinoless double beta decay.
Remarkably, we have found that $s^2_{23}$ value is $s^2_{23} \gtrsim 0.47$ and $s^2_{23} \in [0.55, 0.56]$ near the fixed point of $\tau = \omega$. Moreover, the lightest neutrino mass is $m_1 \sim 50$ meV for $\tau \simeq \omega$ in IH.

We have also checked constraints from LFV decays of charged leptons, $\ell \to \ell' \gamma$. The constraints are easily satisfied when the scale of new physics $M_3$ is sufficiently large. The BRs of cLFV processes are small enough when $M_3 \gtrsim 100$ GeV.

In principle we have dark matter candidate in the model which we assume to be neutral inert scalar boson. Dark matter phenomenology is the same as of the canonical inert Higgs doublet model and it is possible to accommodate the observed relic density via gauge interaction taking into account coannihilation processes and choosing mass parameter. The scalar mass is not very sensitive to neutrino mass and we can also modify singlet fermion mass parameter for fitting. The detailed analysis of dark matter candidate will be considered elsewhere.
\acknowledgments
The order of authors' names is alphabetical. The work was supported by the National Natural Science Fund of China Grant No.~12350410373 (O.~P.) and by the Fundamental Research Funds for the Central Universities (T.~N.).
%
%
\appendix
%

%
%
%
\bibliography{references}
\end{document}